\documentclass[reprint,showpacs,showkeys,amsmath,amssymb,superscriptaddress,aps,prl]{revtex4-1}

\usepackage{graphicx}
\usepackage{epstopdf}
\usepackage{ulem}
\usepackage[dvipsnames]{xcolor}

\begin{document}


\title{Field induced topological Hall effect in non-coplanar triangular antiferromagnetic geometry of Mn$_3$Sn}

\author{Pradeep K. Rout}
\affiliation{School of Physical Sciences, National Institute of Science Education and Research, HBNI, Jatni-752050, India}
\author{P. V. Prakash Madduri}
\affiliation{School of Physical Sciences, National Institute of Science Education and Research, HBNI, Jatni-752050, India}
\author{Subhendu K. Manna}
\affiliation{School of Physical Sciences, National Institute of Science Education and Research, HBNI, Jatni-752050, India}
\author{Ajaya K. Nayak}
\email{ajaya@niser.ac.in}
\affiliation{School of Physical Sciences, National Institute of Science Education and Research, HBNI, Jatni-752050, India}

\date{\today}

\begin{abstract}
Non-collinear triangular antiferromagnets with a coplanar spin arrangement and vanishing net magnetic moment can exhibit a large anomalous Hall effect owing to their non-vanishing momentum space Berry curvature. Here we show the existence of a large field induced topological Hall effect in the non-coplanar triangular antiferromagnetic  geometry of Mn$_3$Sn. A detailed magnetic and Hall effect measurements demonstrate the presence of three distinct Hall contributions; a high temperature anomalous Hall effect, a low temperature toplogical Hall effect and an intermediate temperature region with coexistence of both the effects. The origin of the observed topological Hall effect is attributed to the scalar spin chirality induced real space Berry curvature that appears when the system undergoes from a trivial non-coplanar triangular spin alignement  to a topologically protected non-trivial spin texture like skyrmions by application of magnetic field at low temperatures.

\end{abstract}


\maketitle


Over the past few years, the inclusion of topology in condensed matter research has uncoverd many exotic states of matter, such as topological insulators \cite{Hasan01, Qi02}, topological superconductors \cite{Qi02}, topological Weyl or Dirac fermions \cite{Shun03}, and non-trivial magnetic structures like skyrmions \cite{Pfleiderer04, Yu05, Matsuno06, AKN07, Nagaosa08}. The interplay between the topology and symmetry in some of these materials has led to several novel transport phenomena like anomalous Hall effect (AHE) \cite{AKN09, Nakatsuji10, Liang11} and topological Hall effect (THE) \cite{Matsuno06, Neubauer13, Kanazawa14, Gallagher15, Schulz16, Kanazawa17, Huang18, Li19}. Recent theoretical models have demystified their perplexing origin by invoking the Berry phase concept \cite{Nagaosa20}, which helps in the understanding of recent experimental realization of large AHE in triangular antiferromagnets Mn$_3$Ge and Mn$_3$Sn \cite{AKN09, Nakatsuji10}, frustrated magnetic system Pr$_2$Ir$_2$O$_7$ \cite{Machida21}, and Nd$_2$Mo$_2$O$_7$ spin ice \cite{Taguchi22}. The Berry phase is a topological property of Bloch states and can be defined as a finite geometric phase obtained by the electrons wavefunction when it is adiabatically time evolved in systems with broken time reversal symmetry \cite{Nagaosa20}. With electric field as driving perturbation, where Berry curvature acts like a fictitious emergent magnetic field in K-space, the conduction electrons acquire an anomalous velocity even in zero magnetic field that accounts for the observation of AHE in several magnetic and non-magnetic systems \cite{Karplus24, Onoda25, Nagaosa20}. 


\begin{figure*} [tb!]	
	\includegraphics[angle=0,width=15cm,clip=true]{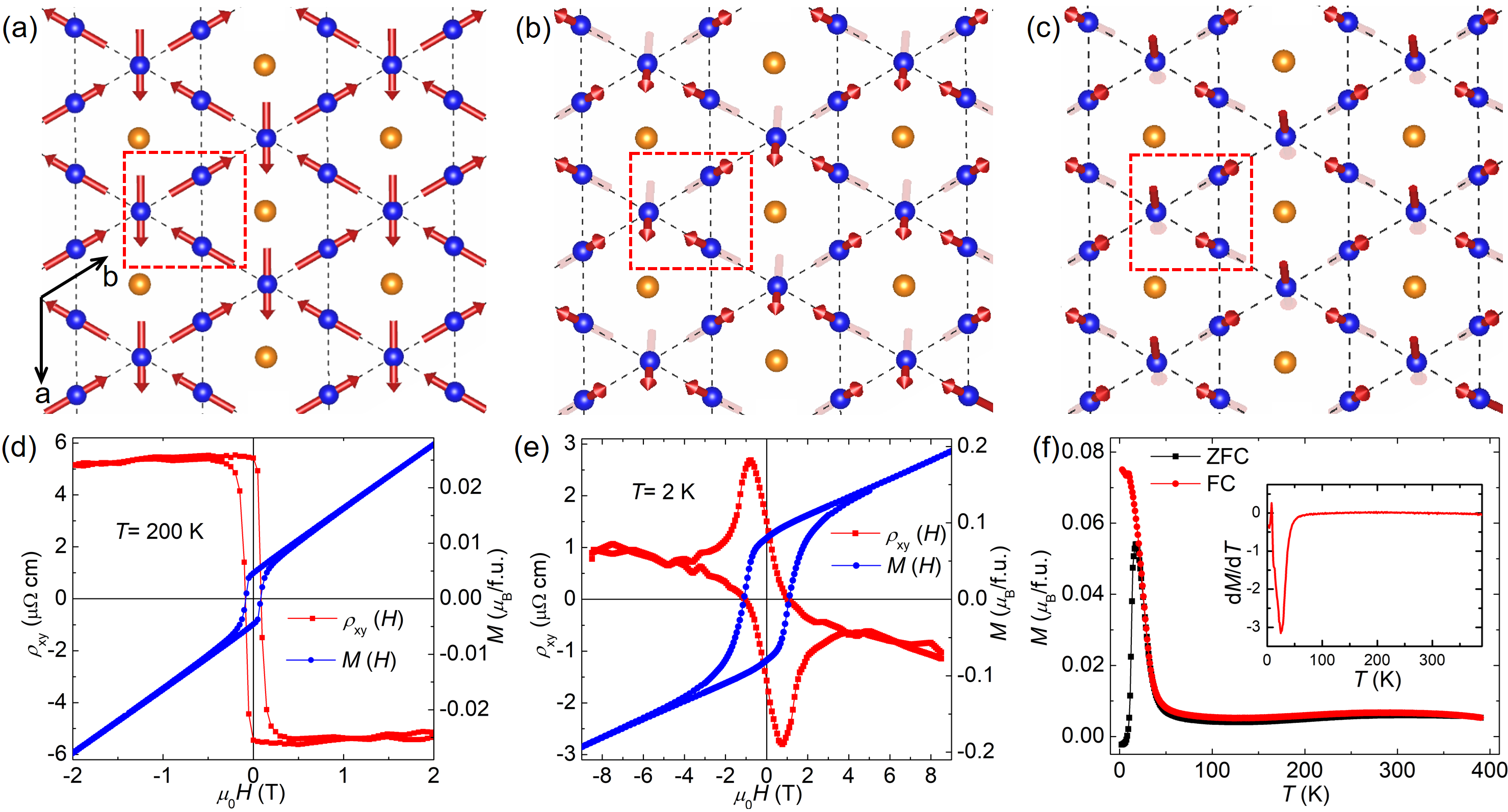}
	\caption{\label{FIG1}(Color online) Atomic arrangement of Mn (blue balls) and Sn (yellow balls) in a single $c$-plane of the hexagonal crystal structure of Mn$_3$Sn.  The magnetic moments of Mn atoms are represented by cylinderical arrows. (a) The 120$^{\circ}$ triangular antiferromagnetic spin alignment of the Mn atoms where all the Mn moments lie in the $c$-plane. (b) All-out configuration,  a slight tilting of the whole spin triangle along the $c$-axis. (c) Two-out-one-in configuration, two Mn moments of the spin triangle tilt along the positive $c$-axis, whereas, one Mn-moment tilts towards the negative $c$-direction. In all cases a single spin triangle is shown inside a dashed box. Field dependence of Hall reistivity ($\rho_{xy}$, left axis) and magnetization ($M$, right axis) measured at (d) 200~K and (e) 2~K. (f) Temperature dependence of magnetization measured in a field of 0.1~T in zero field cooled (ZFC) and field cooled (FC) modes. The inset shows the first derivative of the FC curve.}
\end{figure*}

Topologically non-trivial spin structures having spatial variation in local magnetic moments driven either by geometrical frustration or by Dzylashinsky-Moriya interaction give rise to a different type of Hall effect, termed as topological Hall effect \cite{Onoda25, Bruno26}. The very origin of THE can be attributed to the scalar spin chirality $\chi_{ijk}=\bf S_i \cdot (\bf S_j \times \bf S_k)$, defined as the scalar product corresponds to the solid angle $\Omega$ subtended by the three spins \textbf{S$_i$}, \textbf{S$_j$} and \textbf{S$_k$} on the unit sphere, which breaks the time-reversal symmetry and gives the real space Berry curvature. It should be noted that the net scalar spin chirality in many non-collinear and non-coplanar systems vanishes due to the structural symmetry of the lattices, thereby eliminating the net topological field with the rare exception to special spin structures like kagome, pyrochlore, 4-sublattice ordered triangular lattice, and skyrmion lattice phases \cite{Onoda25, Shiomi29}. The THE originating from the non-zero topological winding of spin textures, such as skyrmions, has been observed in the bulk and thin films of the B20 non-centrosymmetric helimagnets MnSi, FeGe, MnGe \cite{Neubauer13, Kanazawa14, Gallagher15, Schulz16, Kanazawa17, Huang18, Li19}. In these systems the Hall effect corresponds to the Berry curvature in real space and can be understood by the strong coupling regime where the length scale of the spin texture is sufficiently longer than the lattice constant. Although the topological spin vortices, skyrmions, have been mostly found in non-centrosymmetric bulk materials \cite{Bauer27} and chiral thin films \cite{Hoffman28}, there exists a few theoretical elucidations of thier realization in frustrated centrosymmetric materials \cite{Okubo44, Batista42, Leonov43}. Hence, the report of novel topological phenomena pertaining to frustrated centrosymmetric material is still elusive. Chirality driven THE has also been observed in different non-colinear 120$^{\circ}$ spin structures, such as PdCrO$_2$ and Fe$_{1.3}$Sb \cite{Shiomi29, Takatsu30} and non-coplanar antiferromagnetic structure Mn$_5$Si$_3$ \cite{Surgers31}. Observation of spin chirality induced THE in a triangular lattice magnet Fe$_{1.3}$Sb where Dzyaloshinsky-Moriya (DM) interaction modifies the spin structure in the spin clusters associated with interstitial-Fe spins at low temperature is very much interesting \cite{Shiomi29}. Thus, the motivation of finding novel topological phenomena associated with the non-coplanar spin ordering in frustrated kagome lattice antiferromagnets triggered our attention towards readdressing the chiral antiferromagnet Mn$_3$Sn.


The Ni$_3$Sn- type Mn$_3$Sn that crystallizes in a DO$ _{19} $ hexagonal structure with space group P6$_3$/\textit{mmc}, exhibits a chiral 120$^{\circ}$ triangular antiferromagnetic ordering with N\'{e}el temperature of about 420~K \cite{Kouvel33,Tomiyoshi34, Zimmer35, Brown36, Nakatsuji10}. The hexagonal unit cell of  Mn$_3$Sn consists of two sets of Mn triangles stacked along \textit{c}-axis where the Mn atoms form a kagome structure in the $c$-plane and Sn sits at the center of the hexagon [Fig. 1(a)]. Although the neutron diffraction studies by various groups have confirmed the 120$^{\circ}$ triangular spin configuration \cite{Kouvel33, Zimmer35}, the exact orientation of the spin plane and the origin of a weak ferromagnetic behavior are conflicting.  The polarized neutron diffraction study \cite{Tomiyoshi34} performed by Tomiyoshi and Yamaguchi confirmed the existence of an inverse triangular spin structure in the basal plane as depicted in Fig. 1(a). In this structure a small canting of the Mn spins towards their respective easy axes deforms the 120$^{\circ}$ spin triangle resulting in the observed weak ferromagnetism. It has been also proposed that Dzyaloshinskii-Moriya interaction (DMI) with DMI vector parallel to the $c$-axis is necessary to stabilize as well as freely rotate the spin structure in the $c$-plane \cite{Tomiyoshi34, Zimmer35}.

The triangular spin structure of Mn$_3$Sn has led to the observation of  different novel transport phenomena, such as large AHE and large anomalous Nernst effect (ANE), and large magneto-optical Kerr effect (MOKE) \cite{Nakatsuji10, Ikhlas37, Higo38}. Despite the observation of these fascinating properties at room temperature, little attention has been paid towards the magnetic ordering at low temperatures.  Mn$_3$Sn develops a large spontaneous magnetic moment below 50~K \cite{Brown36, Feng39, Tomiyoshi40}. Neutron diffraction studies on single crystal Mn$_3$Sn at low temperatures have suggested a similar triangular antiferromagnetic spin structure with a slight tilting of the spins from the $c$-plane towards the $c$-axis \cite{Tomiyoshi40}. Two different spin canting scenarios have been suggested; (i) all three spins in a triangle tilt towards positive $c$-direction as depcted in Fig. 1(b) and/or (ii) two spins move upwards and the third spin tilts downwards from the $c$-plane [Fig. 1(c)] \cite{Tomiyoshi40}. However, to the best of our knowledge no further study is reported based on the low temperature spin state. In the present study we focus on magnetic and transport measurements to explore the finding of a large topological Hall effect at temperatures below 30~K in polycrystalline Mn$_3$Sn. 

\begin{figure} [tb!]	
	\includegraphics[angle=0,width=8.5cm,clip]{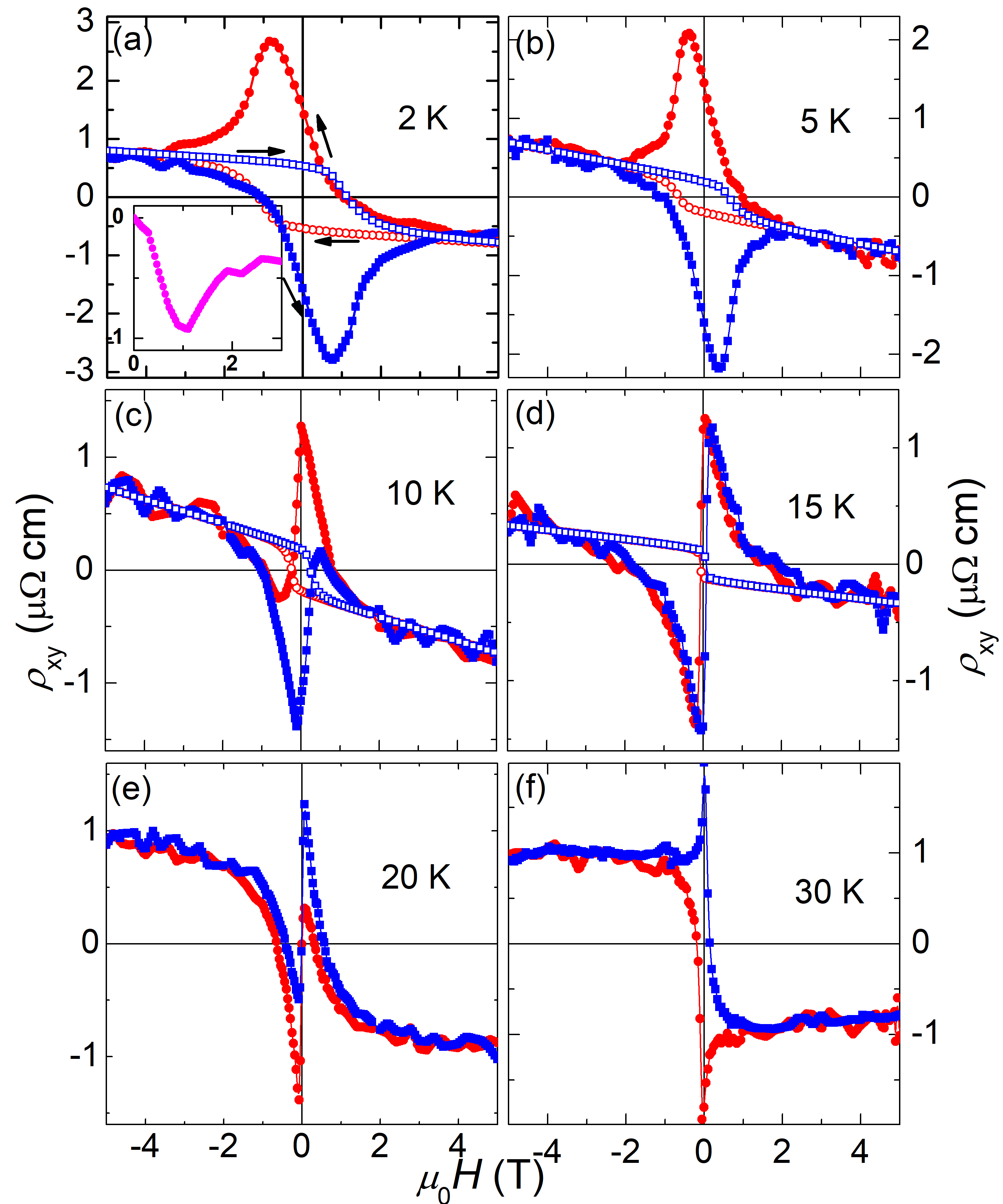}
	\caption{\label{FIG2}(Color online) Field dependence of Hall reistivity [$\rho_{xy}(H)$] measured at different temperatures. The filled circles represent $\rho_{xy}$ data measured from +5~T to -5~T, whereas, $\rho_{xy}$ data from -5~T to +5~T are ploted with  filled squares. The calculated $\rho_{xy}$ curves for $T=2-15$~K are represented by open circles and squares. The inset of (a) represents the virgin curve of $\rho_{xy}(H)$ at 2~K. }
\end{figure}

Polycrystalline ingots of Mn$_{3}$Sn were prepared using arc melting technique. The AHE and the magnetic hysteresis loop measured at 200~K for the present polycrystalline Mn$_3$Sn sample is depicted in Fig. 1(d). A spontaneous anomalous Hall resistivity $\rho_{xy}$ of about 5.5~$\mu\Omega$~cm and spontaneous magnetization of about 0.005~$\mu_\text B$/f.u. are found at 200~K, showing an excellent match with that reported for the single crystalline Mn$_3$Sn \cite{Nakatsuji10}.  $\rho_{xy}(H)$  data measured at higher temperatures also exhibit a similar trend as reported earlier \cite{Nakatsuji10}. As shown in Fig. 1(e), the spontaneous magnetization ($M_\text S$) increases by nearly 20 times to 0.1~$\mu_\text B$/f.u. with a large coercive field ($H_\text C$) of 1~T when measured at 2~K, thus corroborating the previous neutron diffraction study that suggests a small canting of Mn moment along the $c$-direction \cite{Tomiyoshi40}. We have also calculated the tilting angle from the saturation magnetization data at 2~K. The tilting angle for the first case, where all the three spins of the triangle move upwards from $c$-plane to $c$-axis [Fig. 1(b)],  is estimated to be 2.2 degree. Similarly, we also get a tilting angle of 6.5 degree for the alternative spin structure where only two spins move upwards while the third spin tilts downwards from the $c$-plane [see Fig. 1(c)]. The above obtained tilting angles are also in perfect match with the earlier report \cite{Tomiyoshi40}.

 The temperature dependence of magnetization $M(T)$ curves measured  in zero-field-cooled (ZFC) and field-cooled (FC) modes are depicted in Fig. 1 (f). FC magnetization exhibits a sharp rise  below 50~K. A large irreversibility between the ZFC and FC curves below $T$ $\leq$ 25 K indicates  a large out-of-plane magnetic anisotropy. Most importantly, the  $\rho_{xy}(H)$ curve measured at 2~K displays a distinct behavior incomparision to that measured above 50~K. In the following, we will mostly concentrate on the Hall resistivity measurements below 50~K to explore the origin of such peculiar $\rho_{xy}(H)$ behavior.
 

\begin{figure} [tb!]	
	\includegraphics[angle=0,width=8.5cm,clip]{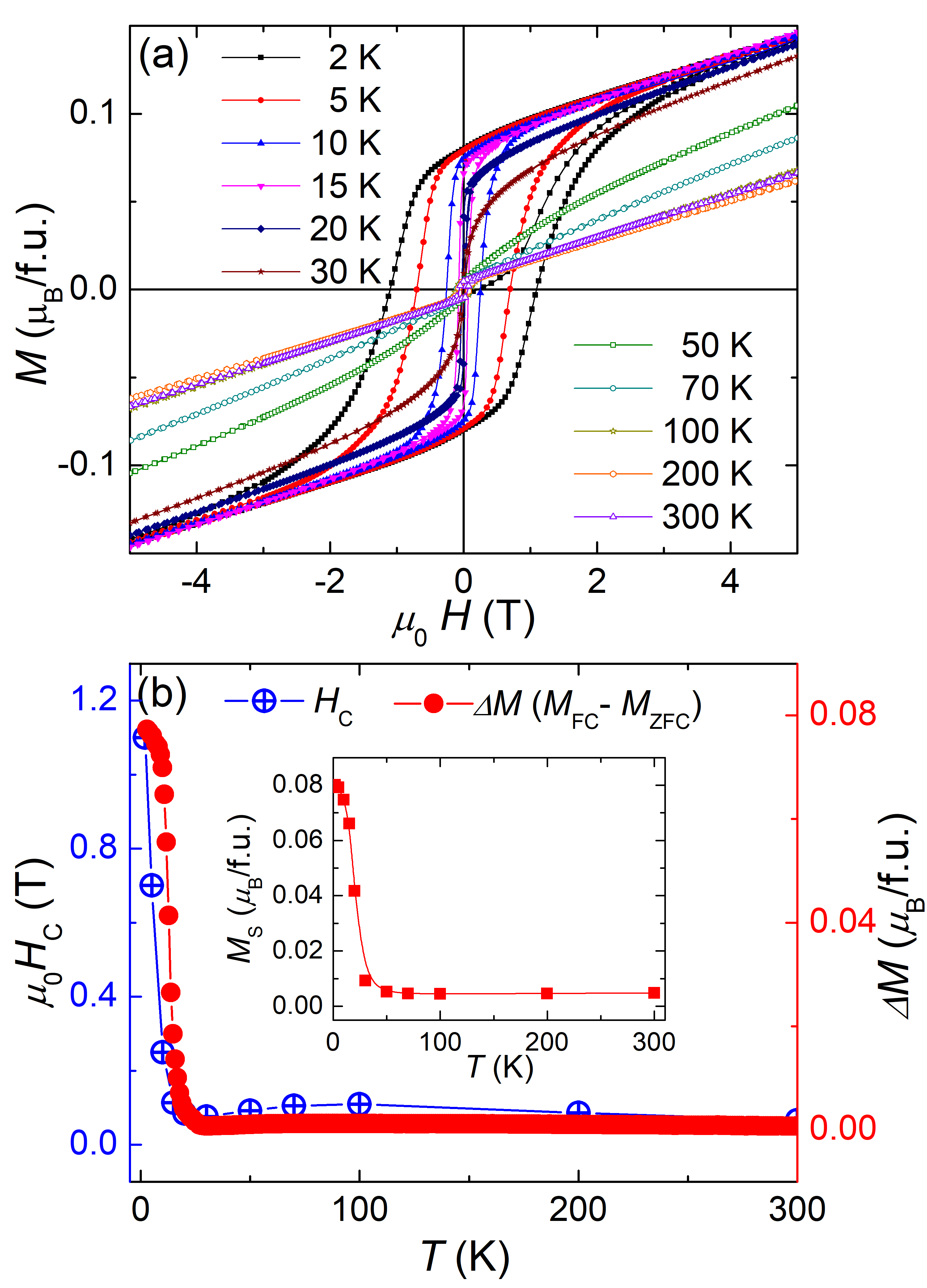}
	\caption{\label{FIG3}(Color online) (a) Field dependence of magnetization [$M(H)$] loops measured at different temperatures. (b) Variation of coercive field [$H_{C}$, left Y-axis] and $\Delta M$ ($M_{FC}-M_{FC}$, right Y-axis) with temperatures. The inset shows the temperature dependence of spontaneous magnetization ($M_S$) derived from the $M(H)$ isotherms.}
\end{figure}

The field dependence of Hall resistivity measured at fixed temperatures in the range 2 ~K - 30 ~K are plotted in Fig. 2 (a) -2 (f). As it can be seen from Fig. 2 (a), the $\rho_{xy}$ measured at 2~K exhibits a positive maxima of 2.7~$\mu\Omega$~cm for field of about - 0.8~T when the field is swept from + 5~T to - 5~T and a negative maxima in the opposite direction. A saturation behavior in the $\rho_{xy}$ data is found for fields above $\pm$~ 3.5 ~T with $\rho_{xy}=0.75$~$\mu\Omega$~cm at  $\pm$ 5~T.  A similar kind of field dependency of the $\rho_{xy}$ data is observed at $T=5$~K. A further increase in temperature to 10~K significantly modifies the $\rho_{xy}$ behavior, where it suddenly falls to negative value  after increasing to a maximum at zero field when the field is swept from + 5~T to - 5~T.  The $\rho_{xy}$ attends a negative value of nearly one sixth of the positive maxima before it again increases to positive saturation. At 15~K, the magnitude of positive maxima and negative maxima of $\rho_{xy}$ around zero field is almost same. However, the positive maximum of $\rho_{xy}$ reduces drastically and the negative maximum increseas to a large value when the field is changed from + 5~T to - 5~T at 20~K. Finally, at 30~K the $\rho_{xy}$  curve measured in + 5~T to - 5~T stays negative up to zero field where it shows a negative peak before changing its sign to positive value in negative field. In all the cases, the observation of peak/hump kind of behavior is only observed when field is swept from positive maximum to negative maximum or vice versa. The observation of peak/hump in the $\rho_{xy}$ data resembles the topological Hall effect behavior that appears with application of magnetic field. This field induced effect can be seen from the virgin $\rho_{xy}$ curve plotted in the inset of Fig. 2 (a), where the peak/hump only appears with application of a magnetic field of nearly 1 T.


\begin{figure} [tb!]	
	\includegraphics[angle=0,width=8.5cm,clip]{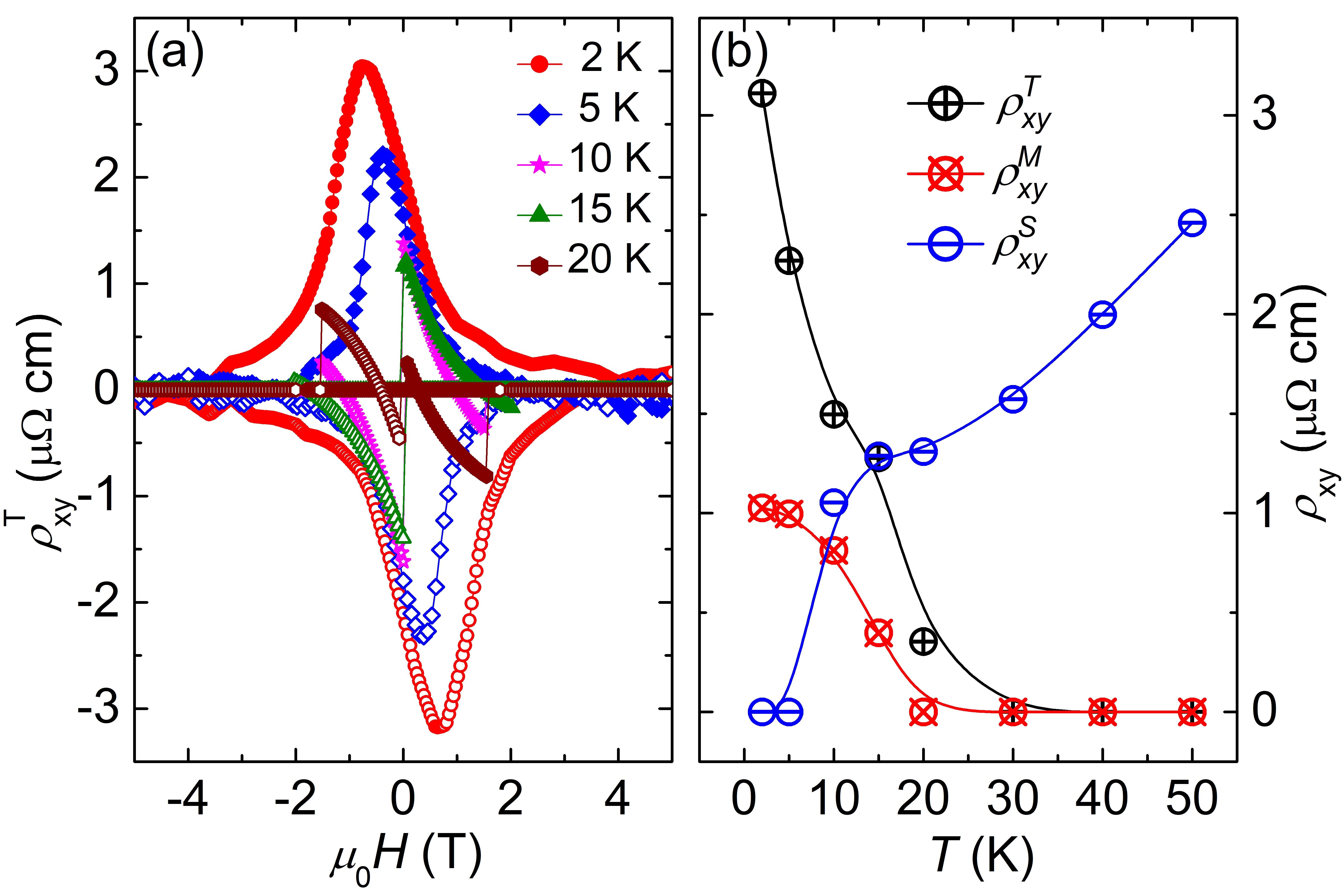}
	\caption{\label{FIG4}(Color online) (a) Field dependence of topological Hall reistivity $\rho^T_{xy}$ at different temperatures. The closed and open symbols represent $\rho_{xy}$  from +5~T to -5~T and -5~T to +5~T, respectively. (b) Maximum value of different components of $\rho_{xy}$ with temperature. $\rho^M_{xy}$ and $\rho^S_{xy}$ are saturated anomalous Hall resistivity scaling with magnetization and spontaneous anomalous Hall resistivity from the coplanar triangular antiferromagnetic structure, respectively. }
\end{figure}

Figure 3 depicts the field dependence of magnetization $M(H)$ curves measured at different temperatures for Mn$_3$Sn.  In contrary to the Hall measurements, no anomaly is found in the $M(H)$ measurements. As expected the coercive field drastically changes from nearly 1~T at 2 K to few Oe's at $T=30$~K. The change in the coercive field is also accompained by a sharp change in the spontaneous magnetization. This indicates that indeed Mn$_3$Sn undergoes a spin reorientaion transition from non-coplanar geometry with a substantial moment along $c$-axis to coplanar antiferromagnetic geometry with all spins lying in the $c$-plane. For the comparison between the observed coercive field and the presence of substantial difference between the ZFC-FC thermomagnetic curve, we have plotted the FC-ZFC magnetization difference ($\Delta M$) and coercive field ($H_C$) in Fig. 3 (b). It can be clearly seen that the $\Delta M$ value remains almost zero down to a temperature of 25 K and sharply increases to 0.08 $\mu_\text B$/f.u. below 25 K. Furthermore, the $H_C$ also follows the same pattern with a steep increase of coercivity below 25 K. The observation of a large coercivity of 1 T at 2 K indicates the presence of large magnetic anisotropy as evident from the $M(T)$ curve in Fig. 1 (f). We have also performed the quantitative estimation of the anisotropy energy from the M (H) loops measured at 2 K and 50 K. The estimated value of the anisotropy energy constant $K_U$ at 2 K is about 2$\times$10$^5$ Jm$^{-3}$. With increasing  temperature, $K_U$ decreases by one order of magnitude to about $K_U$ = 3$\times$10$^4$ Jm$^{-3}$ at 50 K. The spontaneous magnetization plotted in the inset of Fig. 3 (b) mimics temperature dependence behavior of FC magnetization shown in Fig. 1 (f).

\begin{figure} [tb!]	
	\includegraphics[angle=0,width=8cm,clip]{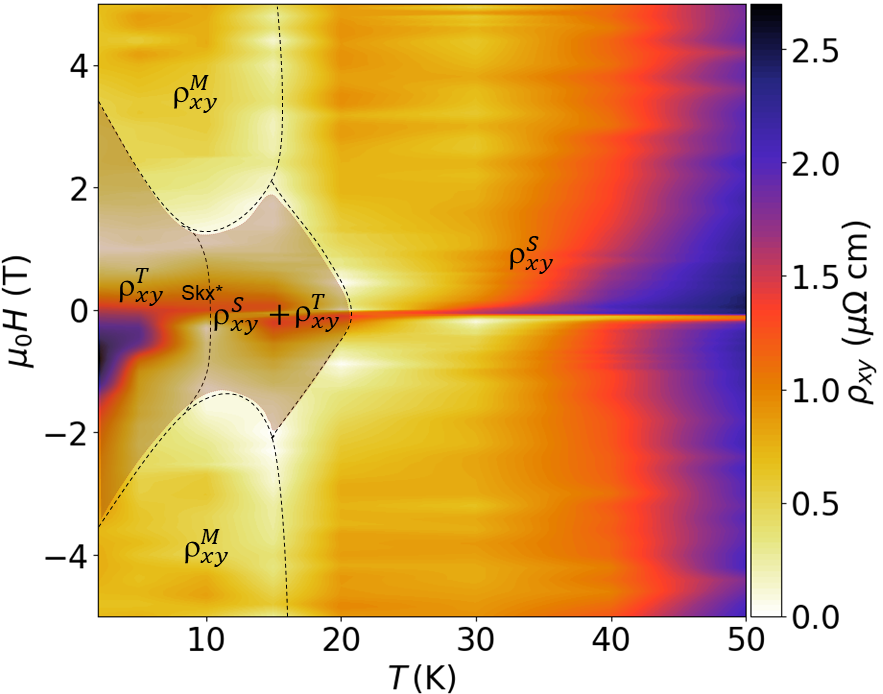}
	\caption{\label{FIG5}(Color online) $H-T$ phase diagram showing contribution from different Hall effects. For the present purpose the  $\rho_{xy}$ measurements with filed sweep from +5~T to -5~T is utilized. Color code represents the magnitude of $\rho_{xy}$ as depicted by scale bar on the right margin. The boundary between different phases are shown by dotted lines. The shaded region marked with Skx$^\star$ represents the possible skyrmion phase. }
\end{figure}

Since the low temperature $M(H)$ isotherms do not exhibit any kind of peculiar behavior, the observed unusual behavior in the $\rho_{xy}$  data  can be understood in terms of unequal contribution from the non-coplanar and coplanar spin textures to the Hall resistivity at different temperatures. It can be mentioned here that the AHE  originating from the triangular coplanar spin structure exhibits a negative saturation value in positive field and vice versa for $T \ge$ 50~K, where the AHE changes its sign with the magnetization due to reversal of spin chirality of the spin triangle \cite{Nakatsuji10}. For $T \le$ 5~K,  the $\rho_{xy}$ curves  change sign independent of the magnetization loop, whereas,  $\rho_{xy}$ changes its sign thrice when the field is swept from + 5~T to - 5~T for 10~K~$ \leq T \leq$ 20~K. In the later case the first sign change of $\rho_{xy}$ curve does not follow the magnetization sign reversal, whereas, the second change of sign around zero field appears at the magnetization switching. Hence for $T \leq$ 20~K, the first sign reversal originates from the field induced topological Hall effect where the electrons scatter in the opposite direction to that of anomolous Hall signal at high fields. The second sign change corresponds to the spontaneous AHE originating from the field independent triangular antiferromagnetic spin structure. Finally, the $\rho_{xy}$ data display a single sign reversal  along with the magnetization loop with a peak kind of behavior around zero field for $T = 30$~K. The above observations suggest that at low temperatures ($T \le$ 5~K) the Hall signal consists of only non-coplanar component, whereas, it is the combination of non-coplanar and coplanar at intermediate temperatures (10~K~$ \leq T \leq$ 20~K) before transfering to the  coplanar one at higher temperatures. For $T \le$ 15~K, the high field ($H > 3$~T) AHE arises from the net uncompensated magnetic moment as the $\rho_{xy}$ scales with the magnetization, i.e, $\rho_{xy}$ weakens with decreasing  magnetization. The high field AHE for $T\geq$ 20 K mostly comes from the coplanar triangular antiferromagnetic spin structure where the field induced non-coplanar effect is minimal.


As it is seen from Fig. 2, the observed  $\rho_{xy}$ behavior can be explained in terms of different Hall components originating from the coplanar and non-coplanar spin structures. The topological Hall contribution extracted from the total  $\rho_{xy}$ is plotted in Fig. 4 (a). At $T= $2~K and 5~K,  the $\rho_{xy}$ data do not consist of Hall contribution from the momentum space Berry curvature emerging from the coplanar triangular antiferromagnetic structure. Hence the total Hall contribution can be expressed as $\rho_{xy} = \rho_{N} + \rho^M_{AH} + \rho^T_{xy}$, where $\rho_{N}$ and $\rho^T_{xy}$ are  normal and topological Hall resistivities, respectively \cite{Neubauer13, Kanazawa14, Gallagher15}. $\rho^M_{AH}$ is the anomalous Hall resistivity proportional to the magnetization in Mn$_3$Sn. Using the relation that the  normal Hall resistivity is proportional to the applied magnetic field and $\rho^M_{xy}$ is proportional to the magnetization, $\rho_{xy}$ curve excluding the topological Hall contribution is calculated and depicted by open symbols in Fig. 2(a) and (b). It can be mentioned here that the present system exhibits a conductivity of the order 10$^{4}$ $\Omega^{-1}$cm$^{-1}$ with almost zero magnetoresistance.  Thus any anomalous Hall contribution originating from the scattering dependent mechanism, which usually dominates in highly conductive systems ($\geq$ 10$^{6}$ $\Omega^{-1}$cm$^{-1}$), can be safely ruled out in our sample \cite{Nagaosa20}. $\rho^T_{xy}$ can be calculated by subtracting the calculated $\rho_{xy}$ data from the measured Hall data, as shown in Fig. 4(a). However, for $T=$ 10~K and 15~K the calculated $\rho_{xy}$ curves shown in Fig. 2(c) and (d), respectively, cannot be used to extract $\rho^T_{xy}$. This is due to the fact that the low field $\rho_{xy}$ data consists of both spontaneous anomalous Hall component $\rho^S_{xy}$ evolving from the coplanar triangular sturcture and the topological Hall component.  As the saturation magnetization decreases the $\rho^M_{xy}$ component ceases to almost zero for $T=$ 20~K and 30~K and $\rho^S_{xy}$ dominates at all fields. Fig. 4(b) shows the temperature dependence of various Hall components. It is clear that the $\rho^S_{xy}$ at any field dominates over all other contributions at temperatures above 20~K.

A $H-T$ phase diagram derived from the field dependence of  $\rho_{xy}$ measurements at different temperatures is displayed in Fig. 5. As discussed above, the maximum contribution of the THE can be found in a small pocket for temperatures less than 10~K and fields between 0 to - 2~T, whereas, a mixed THE and sponatenous Hall effect is obseved between 10~K and 20~K. The observed THE, which appears due to the real space Berry curvature, can be attributed to the formation of non-trivial nanoscale magnetic domains like skyrmions at low temperatures having non-coplanar spin structure.  Since the magnitude of topological Hall voltage is directly proportional to skyrmion density/size, a rough estimation of the size of skyrmion from the relation $\rho^T_{xy}=PR_0B_{eff}$, where $P$ is the conduction electron spin polarization and $B_{eff}$ is the effective (fictitious) magnetic field and $R_0$ is the normal Hall coefficient  \cite{Neubauer13, Kanazawa17, Marrows}, results in a skyrmion diameter of about 1~nm. 


The present finding of large topological Hall effect in the non-coplanar magnetic state is in consistent with  the recent finding of  similar kind of field induced giant topological Hall effect in a frustrated centrosymmetric magnets Gd$_2$PdSi$_3$ and Gd$_3$Ru$_4$Al$_{12}$ with triangular lattice \cite{Kurumaji32, Max}.  The THE below 20~K in Gd$_2$PdSi$_3$ and Gd$_3$Ru$_4$Al$_{12}$ has been attributed to the signature of a skyrmionic state with spin modulation length of about 2~nm \cite{Kurumaji32, Max}. Different mechanisms governing the formation of skyrmions can result in different size of the skyrmionic spin structure. The skyrmion formation due to frustrated exchange interaction can lead to atomic-sized skyrmions of size \textit{a$_{sk}$}$\sim$ 1 nm \cite{Nagaosa08}. These skyrmions exhibit a highly degenerate ground state and have both vorticity as well as helicity degrees of freedom \cite{Nagaosa08}. In general, the smaller sized skyrmions of size 1~nm can squeeze the emergent magnetic flux and produces approximately 4000~T of effective magnetic field which promises  robust storage density while the experimental realization of reading and writing possibilities is still elusive\cite{Kanazawa17}.

Our observation of a large effective magnetic field (\textit{B$_{eff}$}) $\sim$ 3500~T which corresponds to a skyrmion size of $\sim$ 1~nm in the frustrated centrosymmetric Mn$_3$Sn at low temperatures, strongly matches with the above predictions and paves the way towards the realization of skyrmion texture in frustrated magnetic systems. Previous theoretical works have already demonstrated the realization of  metastable multiple periodic states of skyrmions in frustrated centrosymmetric triangular and kagome lattices with easy axis magnetic anisotropy, threefold/sixfold degenerate symmetry axis, and by considering the next nearest neighbor interactions \cite{Batista42, Leonov43, Okubo44}. Frustrated magnetic systems with competing interactions can result in different complex spin structures whose topological nature can be quantified by a geometrical parameter called scalar spin chirality \cite{Machida21, Taguchi22}. Unconventional electromagnetic responses in the form of large unusual Hall effect can also appear in topologically nontrivial spin structure hosting frustrated magnetic materials showing strong electron-electron correlation \cite{Batista42}. Mn$_3$Sn turns out to be a rare example of metallic material exhibiting strong correlation among Mn 3d electrons as observed from the angle resolved photoemission spectroscopy (ARPES) and density functional theory (DFT) calculations \cite{Kuroda45}. Hence the strong electron-electron correlation along with the easy axis magnetic anisotropy in the frustrated kagome lattice magnet Mn$_3$Sn may lead to a transition from topologically trivial magnetic phase to a topologically non-trivial one at low temperatures ($T<20$~K). This may lead to non-zero scalar spin chirality  which results in the appearance of large topological Hall effect  at low temperatures with the proliferation of more topological entities below 10~K.


As mentioned in the introduction the triangular lattice compounds Mn$_5$Si$_3$ and Fe$_{1.3}$Sb also exhibit THE which is attributed to the scalar spin chirality due to the non-collinear spin ordering \cite{Shiomi29, Surgers31}. However, it is to be noted that in case of Mn$_5$Si$_3$ topological Hall effect is observed in a temperature region where there is a modification of crystal structure from centrosymmetric hexagonal  to a non-centrosymmetric orthorhombic one. The authors attribute the observed effect to the presence of complex non-collinear spin texture at low temperatures.  In case for Fe$_{1.3}$Sb the DM interaction modifies the triangular spin structure of the spin clusters associated with the interstitial Fe- spins to generate finite scalar spin chirality as well as the topological Hall effect. The observed THE in Fe$_{1.3}$Sb is explained with invoking a local inversion symmetry breaking between the interstitial Fe spin and triangularly ordered Fe spins. Although a hexagonal  Mn$_3$Sn crystal structure is stabilized only in the off-stoichiometric composition, the excess Mn atoms replace some of the Sn atoms instead of going to the interstitial position, as reported in earlier studies \cite{Tomiyoshi34, Kren, Zhang}. In addition, it is highly unlikely that this small fraction of excess Mn atoms will show a magnetic ordering that will affect the whole spin texture. Thus, the presence of a similar spin state in Mn$_3$Sn to that of Fe$_{1.3}$Sb, which consists of about 33 \% of excess Fe to that of about 1.5 \% of extra Mn in case of Mn$_{3.05}$Sn$_{0.95}$, is highly unexpected.

Also, at low temperatures  Mn$_3$Sn shows a large magnetic ordering behavior with a coercivity of 1 T and spontaneous moment of 0.1 $\mu_\text B$/f.u. in contrast to the complete antiferromagnetic behavior of Mn$_5$Si$_3$ and Fe$_{1.3}$Sb with almost zero spontaneous magnetization. The observed topological Hall resistivity exhibits completely different behavior with distinct bumps/peaks on top of AHE and significant hysteresis in Mn$_3$Sn as compared to  Mn$_5$Si$_3$ and Fe$_{1.3}$Sb where only a curvature like deviation occurs in the linear Hall behavior with no hysteresis.  Again, the low temperature polarized neutron diffraction measurement in Mn$_3$Sn does not propose any modification in the triangular spin structure [14]. According to the earlier reports, for a triangular lattice with 120$^{\circ}$ spin order, the spin chirality cancels out and the topological contribution to the Hall resistivity is usually not expected \cite{Shiomi29, Akagi, Martin}. Thus, there may not be any finite scalar spin chirality originating from the non-collinear/non-coplanar spin structure in Mn$_3$Sn at low temperature. Therefore, it is most likely that the observed THE in Mn$_3$Sn comes from the presence of field induced non-trivial spin structure like skyrmions.

 In summary, we have demonstrated the existence of a large topological Hall effect in the non-coplanar triangular antiferromagnetic geometry of the frustrated kagome lattice system Mn$_3$Sn. The presence of non-coplanar magnetic structure with large uniaxial magnetic anisotropy at low temperatures is established from the magnetic measurements that exhibit a non-zero spontaneous magnetization and a large coercive field below 20~K. The observation of field induced large topological Hall effect indicates that the application of magnetic field transformes the trivial non-collinear spin texture to a topologically protected non-trivial spin state. At very high fields the spin texture again trasforms to a trivial state. We have also established a $H-T$ phase diagram that shows the esistence of various components of Hall effects at different magnetic fields and temperatures. The origin of the present THE is attributed to the presence of topologically protected spin structures like skyrmions with size as small as 1~nm and emergent magnetic field of $\sim$ 3500~T. The small size of the present non-trivial magnetic structures is well within the size of skyrmion texture expected in frustrated magnetic systems predicted by theory. The present work is an important step towards the  understanding of low temperature magnetic states of Mn$_3$Sn that will motivate further study in similar systems.



\begin{acknowledgments}
	
This work was financially supported by Department of Atomic Energy
(DAE) and Department of Science and Technology (DST)-Ramanujan research grant (No. SB/S2/RJN-081/2016) of
the Government of India.

\end{acknowledgments}



\begin{thebibliography}{100}

\bibitem{Hasan01}M. Z. Hasan and C. L. Kane, Rev. Mod. Phys. \textbf{82}, 3045 (2010).

\bibitem{Qi02}X.-L. Qi and S.-C. Zhang, Rev. Mod. Phys. \textbf{83}, 1057 (2011).

\bibitem{Shun03}S.-Q. Shen, Topological Insulators: Dirac Equation in Condensed Matter,  edited by S.-Q. Shen (Springer Singapore, Singapore, 2017), pp. 207–229.

\bibitem{Pfleiderer04} S. M\"{u}hlbauer, B. Binz, F. Jonietz, C. Pfleiderer, A. Rosch, A. Neubauer, R. Georgii, and P. B\"{o}ni, Science {\bf323}, 915 (2009).

\bibitem{Yu05} X. Z. Yu, Y. Onose, N. Kanazawa, J. H. Park, J. H. Han, Y. Matsui, N. Nagaosa, and Y. Tokura, Nature  {\bf465}, 901 (2010).

\bibitem{Matsuno06} J. Matsuno, N. Ogawa, K. Yasuda, F. Kagawa, W. Koshibae,  N. Nagaosa, Y. Tokura, and M. Kawasaki, Sci. Adv. {\bf2}, e1600304 (2016).

\bibitem{AKN07}A. K. Nayak, V. Kumar, T. Ma, P. Werner, E. Pippel, R. Sahoo, F. Damay, U. K. R\"{o}ssler, C. Felser, and S. S. P. Parkin, Nature \textbf{548}, 561 (2017).

\bibitem{Nagaosa08} N. Nagaosa and Y. Tokura, Nature Nanotech. {\bf8}, 899 (2013).

\bibitem{AKN09}A. K. Nayak, J. E. Fischer, Y. Sun, B. Yan, J. Karel, A. C. Komarek, C. Shekhar, N. Kumar, W. Schnelle, J. K\"{u}bler, C. Felser, and S. S. P. Parkin, Sci. Adv. \textbf{2}, e1501870 (2016).

\bibitem{Nakatsuji10}S. Nakatsuji, N. Kiyohara, and T. Higo, Nature \textbf{527}, 212 (2015).

\bibitem{Liang11}T. Liang, J. Lin, Q. Gibson, S. Kushwaha, M. Liu, W. Wang, H. Xiong, J. A. Sobota, M. Hashimoto, P. S. Kirchmann, Z.-X. Shen, R. J. Cava, and N. P. Ong, Nature Phys. \textbf{14}, 451 (2018).


\bibitem{Neubauer13} A. Neubauer, C. Pfleiderer, B. Binz, A. Rosch, R. Ritz, P.G. Niklowitz, and P. B\"{o}ni, Phys. Rev. Lett. {\bf102}, 186602 (2009).

\bibitem{Kanazawa14} N. Kanazawa, Y. Onose, T. Arima, D. Okuyama, K. Ohoyama, S. Wakimoto, K. Kakurai, S. Ishiwata, and Y. Tokura, Phys. Rev. Lett. {\bf106}, 156603 (2011).

\bibitem{Gallagher15}J. C. Gallagher, K. Y. Meng, J. T. Brangham, H. L. Wang, B. D. Esser, D. W. McComb, and F. Y. Yang, Phys. Rev. Lett. {\bf118}, 027201 (2017).

\bibitem{Schulz16} T. Schulz, R. Ritz, A. Bauer, M. Halder, M. Wagner, C. Franz, C. Pfleiderer, K. Everschor, M. Garst, and A. Rosch, Nature Phys. {\bf8}, 301 (2012).

\bibitem{Kanazawa17} N. Kanazawa, M. Kubota, A. Tsukazaki, Y. Kozuka, K. S. Takahashi, and M. Kawasaki, Phys. Rev. B  {\bf91}, 041122(R) (2015).

\bibitem{Huang18} S. X. Huang and C. L. Chien, Phys. Rev. Lett. {\bf108}, 267201 (2012).

\bibitem{Li19} Yufan Li, N. Kanazawa, X. Z. Yu, A. Tsukazaki, M. Kawasaki, M. Ichikawa, X. F. Jin, F. Kagawa, and Y. Tokura, Phys. Rev. Lett. {\bf110}, 117202 (2013).

\bibitem{Nagaosa20}N. Nagaosa, J. Sinova, S. Onoda, A. H. MacDonald, and N. P. Ong, Rev. Mod. Phys. \textbf{82}, 1539 (2010).

\bibitem{Machida21}Y. Machida, S. Nakatsuji, S. Onoda, T. Tayama, and T. Sakakibara, Nature \textbf{463}, 210 (2009).

\bibitem{Taguchi22}Y. Taguchi, Y. Oohara, H. Yoshizawa, N. Nagaosa, and Y. Tokura, Science \textbf{291}, 2573 (2001).

\bibitem{Karplus24}R. Karplus and J. M. Luttinger, Phys. Rev. \textbf{95}, 1154 (1954).

\bibitem{Onoda25}O. Masaru, T. Gen, and N. Naoto, J. Phys. Soc. Japan \textbf{73}, 2624 (2004).

\bibitem{Bruno26} P. Bruno, V.K. Dugaev, M. Taillefumier, Phys. Rev. Lett. \textbf{93}, 096806 (2004)

\bibitem{Bauer27} A. Bauer and C. Pfleiderer, in Topol. Struct. Ferroic Mater. Domain Walls, Vortices and Skyrmions, edited by J. Seidel (Springer International Publishing, Cham, 2016), pp. 1–28.

\bibitem{Hoffman28} W. Jiang, G. Chen, K. Liu, J. Zang, S. G.E. te Velthuis, and A. Hoffmann, Phys. Rep. \textbf{704}, 1 (2017).

\bibitem{Shiomi29}Y. Shiomi, M. Mochizuki, Y. Kaneko, and Y. Tokura, Phys. Rev. Lett. \textbf{108}, 56601 (2012).

\bibitem{Takatsu30}H. Takatsu, S. Yonezawa, S. Fujimoto, and Y. Maeno, Phys. Rev. Lett. \textbf{105}, 137201 (2010).

\bibitem{Surgers31}C. S\"{u}rgers, G. Fischer, P. Winkel, and H. V. L\"{o}hneysen, Nature Commun. \textbf{5}, 6 (2014).

\bibitem{Kurumaji32}T. Kurumaji, T. Nakajima, M. Hirschberger, A. Kikkawa, Y. Yamasaki, H. Sagayama, H. Nakao, Y. Taguchi, T. Arima, and Y. Tokura, ArXiv:1805.10719 (2018).

\bibitem{Kouvel33}J. S. Kouvel and J. S. Kasper,   Proc. Conf. on Magnetism (Nottingham, 1964) (London: Institute of Physics) page 169 (1965).

\bibitem{Tomiyoshi34}T. Shōichi and Y. Yasuo, J. Phys. Soc. Japan \textbf{51}, 2478 (1982).

\bibitem{Zimmer35}G. J. Zimmer and E. Krén, AIP Conf. Proc. \textbf{10}, 1379 (1973).

\bibitem{Brown36}P. J. Brown, V. Nunez, F. Tasset, J. B. Forsyth, and P. Radhakrishna, J. Phys. Condens. Matter \textbf{2}, 9409 (1990).

\bibitem{Ikhlas37}M. Ikhlas, T. Tomita, T. Koretsune, M.-T. Suzuki, D. Nishio-Hamane, R. Arita, Y. Otani, and S. Nakatsuji, Nature Phys. \textbf{13}, 1085 (2017).

\bibitem{Higo38}T. Higo, H. Man, D. B. Gopman, L. Wu, T. Koretsune, O. M. J. vant Erve, Y. P. Kabanov, D. Rees, Y. Li, M.-T. Suzuki, S. Patankar, M. Ikhlas, C. L. Chien, R. Arita, R. D. Shull, J. Orenstein, and S. Nakatsuji, Nature Photonics \textbf{12}, 73 (2018).

\bibitem{Feng39}W. J. Feng, D. Li, W. J. Ren, Y. B. Li, W. F. Li, J. Li, Y. Q. Zhang, and Z. D. Zhang, Phys. Rev. B \textbf{73}, 205105 (2006).

\bibitem{Tomiyoshi40}S. Tomiyoshi, S. Abe, Y. Yamaguchi, H. Yamauchi, and H. Yamamoto, J. Magn. Magn. Mater. \textbf{54}, 1001 (1986).

\bibitem{Marrows} K. Zeissler, S. Finizio, K. Shahbazi, J. Massey, F. Al Ma’Mari, D. M. Bracher, A. Kleibert, M. C. Rosamond, E. H. Linfield, T. A. Moore, J. Raabe, G. Burnell, and C. H. Marrows, Nat. Nanotechnol. \textbf{13}, 1161 (2018).

\bibitem{Batista42} C. D. Batista, S-Z. Lin, S. Hayami, and Y. Kamiya, Rep. Prog. Phys. \textbf{79} 084504 (2016).

\bibitem{Leonov43}A. O. Leonov and M. Mostovoy, Nat. Commun. \textbf{6}, 8275 (2015).

\bibitem{Okubo44}T. Okubo, S. Chung, and H. Kawamura, Phys. Rev. Lett. \textbf{108}, 17206 (2012).

\bibitem{Kuroda45}K. Kuroda, T. Tomita, M.-T. Suzuki, C. Bareille, A. A. Nugroho, P. Goswami, M. Ochi, M. Ikhlas, M. Nakayama, S. Akebi, R. Noguchi, R. Ishii, N. Inami, K. Ono, H. Kumigashira, A. Varykhalov, T. Muro, T. Koretsune, R. Arita, S. Shin, T. Kondo, and S. Nakatsuji, Nature Mater. \textbf{16}, 1090 (2017).

\bibitem{Kren}E. Krén, J. Paitz, G. Zimmer, and É. Zsoldos, Phys. B \textbf{80}, 226 (1975).

\bibitem{Zhang} D. Zhang, B. Yan, S.-C. Wu, J. Kübler, G. Kreiner, S. S. P. Parkin, and C. Felser, J. Phys. Condens. Matter \textbf{25}, 206006 (2013).

\bibitem{Martin} I. Martin and C.D. Batista, Phys. Rev. Lett. \textbf{101}, 156402 (2008).

\bibitem{Akagi} Y.Akagi and Y. Motome, J. Phys. Soc. Jpn. \textbf{79}, 083711 (2010).

\bibitem{Max} M. Hirschberger, T. Nakajima, S. Gao, L. Peng, A. Kikkawa, T. Kurumaji, M. Kriener, Y. Yamasaki, H. Sagayama, H. Nakao, K. Ohishi, K. Kakurai, Y. Taguchi, X. Yu, T. Arima, and Y. Tokura, arXiv:1812.02553 (2018), arXiv: 1812.02553.





\end{thebibliography}
\end{document}